Glycerol Inhibits Water Permeation through *Plasmodium Falciparum* Aquaglyceroporin

Liao Y. Chen[1][†]

[1]Department of Physics, University of Texas at San Antonio, One UTSA Circle, San Antonio, Texas 78249 USA

Glycerol modulates *P. Falciparum* Aquaglyceroporin

[†]To whom correspondence should be addressed:

Liao Y. Chen

Department of Physics

One UTSA Circle, San Antonio, Texas 78249, USA

Tel: (210)458-5457

Fax: (210)458-4919

Email: Liao.Chen@utsa.edu

## ABSTRACT

*Plasmodium falciparum* aquaglyceroporin (PfAQP) is a multifunctional membrane protein in the plasma membrane of *P. falciparum*, the parasite that causes the most severe form of malaria. The current literature has established the science of PfAQP's structure, functions, and hydrogen-bonding interactions but left unanswered the following fundamental question: Does glycerol modulate water permeation through aquaglyceroporin that conducts both glycerol and water? This paper provides an affirmative answer to this question of essential importance to the protein's functions. On the basis of the chemical-potential profile of glycerol from the extracellular bulk region, throughout PfAQP's conducting channel, to the cytoplasmic bulk region, this study shows the existence of a bound state of glycerol inside aquaglyceroporin's permeation pore, from which the dissociation constant is approximately 14 μM. A glycerol molecule occupying the bound state occludes the conducting pore through which permeating molecules line up in single file by hydrogen-bonding with one another and with the luminal residues of aquaglyceroporin. In this way, glycerol inhibits permeation of water and other permeants through aquaglyceroporin. The biological implications of this theory are discussed and shown to agree with the existent *in vitro* data. It turns out that the structure of aquaglyceroporin is perfect for the van der Waals interactions between the protein and glycerol to cause the existence of the bound state deep inside the conducting pore and, thus, to play an unexpected but significant role in aquaglyceroporin's functions.

**INTRODUCTION**

*Plasmodium falciparum* aquaglyceroporin (PfAQP) is a member of the membrane channel proteins responsible for transport of water, glycerol, and some other selected solutes across the cell membrane.(Agre, 2008; Agre et al., 1998; Borgnia et al., 1999; Carbrey and Agre, 2009; Heymann and Engel, 1999; Lee et al., 2004) The science is clear that PfAQP conducts both water and glycerol and that the amphipathic pore of PfAQP selectively facilitates the passage of waters and glycerols through the conducting channel by hydrogen-bonding interactions between the luminal residues and the permeants as well as among the permeants.(Beitz, 2007; Beitz et al., 2004; Newby et al., 2008; Schnick et al., 2009) However, a question remains unanswered: Does glycerol modulate or inhibit water permeation through PfAQP channel that conducts both glycerol and water? This fundamental question can be resolved once we have an accurate determination of the chemical potential of glycerol as a function of its center-of-mass coordinates along a path leading from the periplasm to the entry vestibule of PfAQP, through the channel, to the cytoplasm. This chemical-potential profile, considered on the basis of the structure information available in the literature,(Newby et al., 2008) can ascertain the conclusion that glycerol inhibits water permeation through PfAQP.

Inside the PfAQP channel, waters and glycerols line up in a single file, occluding one another from occupying the same z-coordinate. (The z-axis is chosen as normal to the membrane-water interface, pointing from the periplasm to the cytoplasm.) Therefore, waters and glycerols permeate through the amphipathic pore of PfAQP in a concerted, collective diffusion. If a deep enough chemical-potential well exists inside the channel (where the chemical potential is lower than the periplasm/cytoplasm bulk level), a glycerol molecule will be bound there, with a probability determined by the glycerol concentration and the dissociation constant. The bound glycerol will occlude permeation of other molecules through the channel. PfAQP will switch between being open and closed to water permeation as a glycerol is dissociated from and bound to the binding site inside the protein's conducting pore. Therefore, such a chemical-potential profile means that glycerol modulates water permeation through PfAQP in the glycerol concentration range around the dissociation constant. And the half-maximal inhibitory concentration ($IC_{50}$) is approximately equal to the dissociation constant.

In order to produce an accurate estimate of the chemical-potential profile of glycerol, I built an all-atom model of the PfAQP-membrane system using the CHARMM 36 force fields.(Brooks et al., 2009; MacKerell et al., 2000) I conducted a total of 761 ns equilibrium molecular dynamics (MD) and non-equilibrium steered molecular dynamics (SMD) simulations, which amount to approximately an order of magnitude more than the computing efforts invested on any one aquaglyceroporin in a published work of the current literature.

The main results are shown in Fig. 1. Glycerol is found to have a deep chemical-potential well in the PfAQP channel near the Asn-Leu-Ala (NLA) and Asn-Pro-Ser (NPS) motifs that is 6.5 kcal/mol below its chemical potential in the bulk of periplasm/cytoplasm, which corresponds to a dissociation constant of 14 µM. Glycerol binding to or dissociating from this binding site strongly modulates water permeation through the PfAQP pore with an $IC_{50}$ of 14 µM. This theory of glycerol inhibiting PfAQP fully agrees with the existing *in vitro* data. Its validation can be furthered by future experiments measuring the glycerol-PfAQP dissociation constant and the water permeability in the full range of glycerol concentration from µM to mM.

**METHODS**

**System setup.** This study was based on the following all-atom model of PfAQP in the cell membrane (Fig. 2): The PfAQP tetramer, formed from the crystal structure (PDB code: 3C02) with 20 glycerols, was embedded in a patch of fully hydrated palmitoyloleylphosphatidyl-ethanolamine (POPE) bilayer. The PfAQP-POPE complex is sandwiched by two layers of water, each of which is approximately 40Å in thickness. The system is neutralized and ionized with $Na^+$ and $Cl^-$ ions at a concentration of 150 mM. The entire system, consisting of 150,459 atoms, is 116Å×114 Å×107 Å in dimension when fully equilibrated. This system (SysI) has a glycerol concentration of 24 mM. A second system (SysII), for 0 M glycerol, was derived from SysI by deleting all 20 glycerols. It has 150,179 atoms in all and dimensions approximately equal to those of SysI. The Cartesian coordinates are chosen such that the origin is at the geometric center of the PfAQP tetramer. The xy-plane is parallel to the lipid-water interface and the z–axis is pointing from the periplasm to the cytoplasm.

All the simulations of this work were performed using NAMD 2.8.(Phillips et al., 2005) The all-atom CHARMM36 parameters(Brooks et al., 2009; MacKerell et al., 2000) were adopted for all the inter- and intra-molecular interactions. Water was represented explicitly with the TIP3 model. The pressure and the temperature were maintained at 1 bar and 293.15 K, respectively. The Langevin damping coefficient was chosen to be 5/ps. The periodic boundary conditions were applied to all three dimensions, and the particle mesh Ewald was used for the long-range electrostatic interactions. Covalent bonds of hydrogens were fixed to their equilibrium length. The time step of 1 fs was used for short-range interactions and the time step of 4 fs was used for long-range forces in both equilibrium and nonequilibrium simulations. The cut-off for long-range interactions was set to 12Å with a switching distance of 10Å. In all simulations, the alpha carbons on the trans-membrane helices of PfAQP within the range of $-10Å<z<10Å$ were fixed to fully respect the crystal structure.

**Equilibrium MD.** Two runs of 100 ns each in length were conducted for SysI and SysII respectively. Using the theoretical formulation of Zhu *et al*,(Zhu et al., 2004) I computed the mean square displacements (MSD) of the water molecules in the conducting pore $<\delta n^2(t)>$, as defined in Refs. (Jensen and Mouritsen, 2006; Zhu et al., 2004). The osmotic permeation rate of water, $p_f=V_w D_n$, is related to the slope of an MSD curve, $D_n=<\delta n^2(t)>/2t$.(Jensen and Mouritsen, 2006; Zhu et al., 2004) Here $V_w=3\times10^{-23} cm^3$ is the average volume occupied by one water molecule.

**Non-equilibrium SMD.** SMD runs were conducted to sample the transition paths of glycerol going from the periplasm, through the conducting pore, to the cytoplasm, for computing the chemical-potential of glycerol as a function of its center-of-mass position**.** Two sets of SMD simulations were completed to achieve reliable statistics. In each set of SMD, the starting structure is the fully equilibrated structure of SysI. The center of mass of a glycerol was steered/pulled in the positive z-direction to sample a forward pulling path, and then pulled in the negative z-direction to sample a reverse pulling path. The entire path leading from the periplasm (z<−25Å), through the PfAQP pore, to the cytoplasm bulk region (z>25Å), was divided into 50 sections of 1.0 Å each in width. In Set #1, five forward paths and five reverse paths were sampled in each of the 50 sections. In Set #2, four forward paths and four reverse paths were sampled in each of the 22 sections of 1.0 Å each in width between (−10Å<z<12Å) covering the single-file region of the conducting channel where waters and glycerols line up in a single file. In both sets, 4.0 ns equilibration was performed at both end points of each section so that the pulling paths were sampled between equilibrium states with the glycerol's center-of-mass coordinates being fixed at the desired values. And, in all cases, the pulling speed was v=2.5 Å /ns.

Along each forward pulling path from $A$ to $B$, the work done to system was recorded as $W_{A \to Z}$ when glycerol was pulled from $A$ to $Z$. Along each reverse pulling path from $B$ to $A$, the work done to system was recorded as $W_{B \to Z}$ when glycerol was pulled from B to Z. Here $Z$ represents a state of the system when the center-of-mass z-coordinate of the pulled glycerol is $z$. A and B represent two end states of a given section, respectively. The chemical potential (the standard free energy(Serdyuk, 2007)) of glycerol, $\mu(x,y,z)$, a function of its center-of-mass position along the permeation path, can be computed from the Brownian dynamics fluctuation-dissipation theorem (BD-FDT) as follows (Chen, 2011; Chen et al., 2010) :

$$\mu(x, y, z) - \mu(x_A, y_A, z_A) = -k_B T \ln \left( \frac{\langle \exp[-W_{A \to Z} / 2k_B T] \rangle_F}{\langle \exp[-W_{Z \to A} / 2k_B T] \rangle_R} \right). \tag{1}$$

Here $W_{Z \to A} = W_{B \to A} - W_{B \to Z}$ is the work done to the system for the part of a reverse path when the glycerol was pulled from Z to A. $k_B$ is the Boltzmann constant and $T$ is the absolute temperature. $z_A$ and $z_B$ are the z-coordinates of the center of mass of the pulled glycerol at the end states $A$ and $B$ of the system, respectively. From the chemical potential, one can approximate the potential of mean force (PMF) as a function of only the z-coordinate:

$$G(z) = -k_B T \ln \left( \int dxdy \exp[-\mu(x,y,z)/k_B T] / A_{\text{ref.}} \right)$$
$$\simeq \mu(x^*(z), y^*(z), z) - k_B T \ln(A(z)/A_{\text{ref.}}) \qquad (2)$$

where $A_{\text{ref.}}$ and $A(z)$ are, respectively, the area for reference and the area occupied by the center of mass of glycerol on the plane of a given z-coordinate. They are directly related to the glycerol and protein concentrations. $x^*(z)$ and $y^*(z)$ are the median coordinates of integration on the same plane. The second term of the second line of Eq. (2) is the entropic penalty. This term is proportional to the temperature. It does modify the overall rate of transport but does not alter the Arrhenius activation barrier.

**RESULTS AND DISCUSSION**

**Chemical-potential profile of glycerol.** Shown in Fig. 1 is the chemical potential of glycerol as a function of its center-of-mass position along a path leading from the periplasm to the entry vestibule of PfAQP, through the channel, to the cytoplasm. Inside the channel, waters and glycerols line up in a single file, occluding one another from occupying the same z-coordinate. Outside the channel, farther away from the protein, there is more and more space for multiple waters and glycerols to occupy the same z-coordinate. In the single-file region ($-10\text{Å} < z < 12\text{Å}$), the chemical-potential is along the most probable path because the z-coordinate of the glycerol's center of mass was steered while the x- and y- coordinates were freely determined by the system's dynamics. In the non-single-file regions, the chemical potential curve in Fig. 1 is along two straight lines leading from the periplasm to the channel entrance and from the channel exit to the cytoplasm.

It needs to be noted here that chemical potential is not exactly identical to the potential of mean force (PMF) used in, *e.g.*, Refs. (Aponte-Santamaria et al., 2010; Hub and de Groot, 2008). (See Eq. (2) in the previous section.) In the non-single-file regions outside the conducting pore, there are many (infinite) possible paths for a glycerol to dissociate from PfAQP. The PMF, being a function of only the z-coordinate of the glycerol's center of mass, necessarily involves the information of the glycerol and protein concentrations. In contrast, the chemical potential is computed here as a function of the glycerol's center-of-mass position (x-, y-, and z-coordinates) along two chosen straight lines leading from the protein to the periplasm and to the cytoplasm, respectively. It does not incorporate the concentration-dependence (namely, the entropic contributions) here. Considering this chemical potential (the standard free energy) along with the concentration terms,(Chandler, 1987; Serdyuk, 2007) one can show that the absolute binding free-energy of glycerol is equal to the chemical-potential difference between the apo state ($z=25$Å) and the bound state ($z=3$Å): $E_b=-6.5\pm1.5$ kcal/mol.

Correspondingly, the dissociation constant of glycerol from PfAQP is $k_d=\exp[E_b/k_BT]\sim 14$ µM. It should be emphasized that the binding site is in the single-file region. Therefore, water permeation through PfAQP is inhibited by glycerol dwelling at the binding site because a glycerol bound inside PfAQP occludes the entire conducting channel. The half-maximal inhibitory concentration is equal to the dissociation constant, $IC_{50}=k_d$. Currently, there is no means to conduct a direct experimental confirmation of the chemical-potential curve in Fig. 1 but it is feasible to verify the consequences of such a chemical-potential profile. The following biophysical implications of the theory are verifiable by *in vitro* experiments and, indeed, the first two are validated with the existent *in vitro* data in the current literature.

*1. Binding sites of glycerol in PfAQP's pore.* The chemical-potential landscape (Fig. 1, top panel) corresponds well with the pore radius of PfAQP along the channel (Fig. 3). Where there is a maximum in pore radius, there is a minimum in chemical potential. There are two chemical-potential minima in the single-file region---one near the selectivity-filter (SF) and one in the region of the NLA-NPS motifs. The locations of these minima are in agreement of the structural studies of PfAQP that show a glycerol near the SF and another near the NLA-NPS motifs (Fig. 1, bottom panel).(Newby et al., 2008) The two minima are both far below the bulk chemical-potential level. A glycerol bound at either of the two minima will occlude the conducting channel of PfAQP.

*2. Rates of water permeation through PfAQP vs glycerol concentration.* The chemical-potential profile in Fig. 1 implies strong inhibition of water transport by glycerol because a glycerol molecule dwelling in the chemical-potential well occludes the conducting pore. The $IC_{50}$ is approximately 14 µM. In another word, the water permeability of PfAQP in absence of glycerol should be significantly higher than the case with practically any amount of glycerol present. This prediction is in perfect agreement with the existing *in vitro* experiments:

In (Hansen et al., 2002), Hansen *et al* demonstrated the water permeability of PfAQP is nearly equal to that of the water-selective AQP1 (276 *versus* 304 µm/s) when glycerol was not used during the water transport assays. Both of those permeabilities are more than 16 times higher than the water permeability of control Xenopus oocytes (17 µm/s).

In a 2012 *in vitro* study,(Song et al., 2012) Song *et al* examined water permeability of PfAQP expressed in protoplasts. They determined the rate constants from the exponential fitting of their light scattering curves. The rate constants so determined are proportional to the water permeation rates across the protoplasts. They extracted the difference between the rate of PfAQP-expressing protoplasts and the rate of the non-expressing control protoplasts for different osmolytes including glycerol. For 300 mosM across the protoplasts induced by sucrose in absence of glycerol, the rate difference was 0.3/s. For the same 300 mosM induced by glycerol, the rate difference was only 0.01/s, indicating the PfAQP was completely inhibited at high glycerol concentration.

Interestingly, in the transport assays of *(Newby et al., 2008),* glycerol was not added to the assay buffers for water permeation measurements. But glycerol was used throughout the protein preparation procedures. The purified proteins were saturated with glycerols due to the high glycerol concentration in the preparation buffers. Those glycerols were not removed from the proteins but brought into the transport assay buffers. This gave a low but nonzero glycerol concentration (an estimation would give us 12 µM) in the water transport assay buffer. Rates of water conductance at 4°C were found to be 21.5±0.8 for proteoliposomes *vs* 4.5±0.1 for control liposomes, an approximately five-fold increase in water permeation by PfAQP reconstitution into the liposomes. In comparison with this five-fold increase, the increase due to AQP1 reconstitution was found(Zeidel et al., 1992) to be 76 folds at 5°C (extrapolated from measured data in the temperature range from 8.5°C to 37°C). And the increase due to AQPZ reconstitution was approximately 64 folds at 5°C.(Borgnia and Agre, 2001) Put these *in vitro* data together, it is clear that PfAQP was permeable to water in the experiments of Newby *et al,(Newby et al., 2008)* but it was less permeable than AQP1 or AQPZ, indicating partial inhibition of water permeation through PfAQP by glycerol in the μM range.

To make this aspect explicit, I conducted two equilibrium MD simulations of 100 ns each with 24 mM (SysI) and 0 mM (SysII) glycerol concentrations, respectively. I computed the rates of water permeation through PfAQP from the mean square displacements of waters inside the conducting pore of PfAQP (shown in Fig. 4).(Zhu et al., 2004) The osmotic permeability of water at 0 mM glycerol concentration was computed to be $6.5\pm0.6$ cm$^3$/s. And the osmotic permeability at 24 mM glycerol concentration was estimated to be $0.7\pm0.6$ cm$^3$/s. These rate estimates and the afore-listed *in vitro* measurements together point to the conclusion one can draw from the chemical-potential in Fig. 1 that glycerol inhibits water transport through PfAQP with an IC$_{50}$ below the mM range. Further functional experiments in this concentration range around the IC$_{50}$ are expected to demonstrate significant variation in the water permeation rate from the glycerol-free limit to the glycerol-saturated limit.

*3. Activation barrier for glycerol permeation.* The highest chemical-potential barrier is at the SF that is $11.5\pm1.2$ kcal/mol above the bottom of the chemical-potential well near the NLA-NPS. This means that glycerol transport has an Arrhenius activation barrier approximately 11.5 kcal/mol in similarity to GlpF (Chen, 2012). This can be validated by *in vitro* experiments measuring glycerol permeation rates at multiple temperatures in a way similar to what was done for GlpF.(Borgnia and Agre, 2001)

*4. Activation barrier for water permeation.* In the bound state, a glycerol molecule resides in the single-file region, deep inside PfAQP's conducting channel, occluding waters or other glycerols from traversing the channel. When the glycerol concentration is in the mM range (far above the dissociation constant, 14 μM), the PfAQP pore is practically saturated with glycerol. Water permeation is fully correlated with the glycerol dissociation. Therefore, the Arrhenius barrier for water permeation through PfAQP at mM glycerol concentrations is no less than the activation barrier for glycerol dissociation to the cytoplasm (8.5 kcal/mol). This theoretical result of the activation barrier can be verified from measurements of the water permeation rates at multiple temperatures. At 0 M glycerol concentration, PfAQP's conducting pore will be glycerol-free. The chemical-potential of water permeation is expected to resemble GlpF, being flat throughout the open channel, having a low Arrhenius barrier, $E_a^0$<4 kcal/mol,(Aponte-Santamaria et al., 2010; Chen et al., 2010; de Groot and Grubmüller, 2001) which is similar to that of AQPZ.(Pohl et al., 2001)

**Relevant interactions.** Now, what interactions are responsible for the chemical-potential profile shown in Fig. 1? Full understanding has been achieved in regards to the hydrogen-bonding of glycerol/water to the luminal atoms of PfAQP throughout its amphipathic pore.(Newby et al., 2008) These interactions do not give rise to high barriers or deep wells because the number of hydrogen bonds that a glycerol/water can form with PfAQP does not vary drastically with its location throughout the conducting channel. However, the dimension of the conducting pore is not uniform (Fig. 3). The narrowest part of the channel is in the SF region, and another narrow part is between the NLA-NPS motifs and the channel exit on the cytoplasmic side. In the region of the NLA-NPS motifs, the channel is still single-file in nature but wider than the two narrow regions. Interestingly enough, even the narrowest part of the PfAQP channel (3.3 Å in diameter) is wider than the diameter of a water molecule (2Å). Water can traverse the entire channel without causing structural distortions to the protein. Permeation of water, in the absence of glycerols, is dominated by the hydrogen-bonding interactions among waters and with the luminal atoms. In contrast, the size of a glycerol molecule is 3.6Å in the least extended dimension. It cannot traverse the PfAQP pore as freely as a water molecule does. In fact, it is the close contact between a glycerol molecule and the PfAQP lumen that gives rise to the chemical-potential profile shown in Fig. 1. The width of the pore turns out to be such that the van der Waals (VDW) forces on the glycerol by the surrounding residues are all attractive in the NLA-NPS region, causing the chemical potential there to be lower than the bulk level. The bound state of a glycerol deep inside the PfAQP channel owes its existence to the attractive VDW interactions between glycerol and the residues near the NLA-NPS motifs. The two chemical-potential barriers result mainly from the repulsive VDW forces on the glycerol by the residues in the two narrow parts of the PfAQP channel (Fig. 5, left panel). The barrier at the SF also has significant contributions from the conformational changes (Fig 5, right panel) of the SF-forming residues caused by the presence of the glycerol.

**Similarity between PfAQP and GlpF**. *Escherichia coli* aquaglyceroporin (GlpF) is the other aquaglyceroporin whose structure and functions were well established in the literature. In (Chen, 2012), I presented the *in silico* study of GlpF. The chemical-potential profile of glycerol in GlpF closely resembles that of glycerol in PfAQP. The implications of glycerol inhibiting water permeation through GlpF are similar to PfAQP and are consistent with the existing *in vitro* data as well.

**CONCLUSION**

Based on the existent *in vitro* experiments and the chemical-potential profile mapped out in the present study, it is logical to suggest that glycerol bound inside PfAQP's conducting channel inhibits water permeation through its amphipathic pore. As an addition to the science of hydrogen-bonding of waters and glycerols in the conducting pore, this study demonstrates that the van der Waals interactions between PfAQP and a glycerol play a distinctive biological role. The size of the conducting pore is such that a region exists near the NLA-NPS motifs where the VDW interactions between PfAQP and a glycerol are attractive. This precise steric arrangement of PfAQP causes the glycerol's chemical potential there to be lower than its bulk level and, therefore, a bound state of glycerol exists deep inside the single-file channel. Finally, this theoretical study can have very serious medical implications. It leads to this hypothesis: When the glycerol concentration is much higher than the dissociation constant (14 µM), P. falciparum will lose its abilities to maintain hydro-homeostasis and to excrete metabolic wastes through PfAQP across its plasma membrane. If indeed P. falciparum has no other means than PfAQP for water transport and for waste excretion,(Beitz, 2007) the parasite will be killed by oversupply of its nutrient (glycerol)!

**ACKNOWLEDGEMENTS**

The author acknowledges support from the NIH (Grant #GM084834) and the Texas Advanced Computing Center.

References

Agre, P., 2008. Aquaporin water channels: From atomic structure to clinical medicine. Neurosci. Res. 61, S2-S2.
Agre, P., Bonhivers, M., Borgnia, M.J., 1998. The aquaporins, blueprints for cellular plumbing systems. J Biol Chem 273, 14659-14662.
Aponte-Santamaria, C., Hub, J.S., de Groot, B.L., 2010. Dynamics and energetics of solute permeation through the Plasmodium falciparum aquaglyceroporin. Phys Chem Chem Phys 12, 10246-10254.
Beitz, E., 2007. Jammed traffic impedes parasite growth. Proceedings of the National Academy of Sciences 104, 13855-13856.
Beitz, E., Pavlovic-Djuranovic, S., Yasui, M., Agre, P., Schultz, J.E., 2004. Molecular dissection of water and glycerol permeability of the aquaglyceroporin from Plasmodium falciparum by mutational analysis. Proc. Natl. Acad. Sci. USA 101, 1153-1158.
Borgnia, M., Nielsen, S., Engel, A., Agre, P., 1999. Cellular and molecular biology of the aquaporin water channels. Annu Rev Biochem 68, 425-458.
Borgnia, M.J., Agre, P., 2001. Reconstitution and functional comparison of purified GlpF and AqpZ, the glycerol and water channels from Escherichia coli. Proc Natl Acad Sci U S A 98, 2888-2893.
Brooks, B.R., Brooks, C.L., Mackerell, A.D., Nilsson, L., Petrella, R.J., Roux, B., Won, Y., Archontis, G., Bartels, C., Boresch, S., Caflisch, A., Caves, L., Cui, Q., Dinner, A.R., Feig, M., Fischer, S., Gao, J., Hodoscek, M., Im, W., Kuczera, K., Lazaridis, T., Ma, J., Ovchinnikov, V., Paci, E., Pastor, R.W., Post, C.B., Pu, J.Z., Schaefer, M., Tidor, B., Venable, R.M., Woodcock, H.L., Wu, X., Yang, W., York, D.M., Karplus, M., 2009. CHARMM: The biomolecular simulation program. Journal of Computational Chemistry 30, 1545-1614.
Carbrey, J.M., Agre, P., 2009. Discovery of the Aquaporins and Development of the Field of Aquaporins, p. 3-28, in: E. Beitz, (Ed.), Springer Berlin Heidelberg.
Chandler, D., 1987. Introduction to Modern Statistical Mechanics Oxford University Press, Oxford.
Chen, L.Y., 2011. Exploring the free-energy landscapes of biological systems with steered molecular dynamics. Physical Chemistry Chemical Physics 13, 6176-6183.
Chen, L.Y. 2012. Glycerol Modulates Water Permeation through Escherichia coli Aquaglyceroporin GlpF, http://arxiv.org/abs/1209.3264.
Chen, L.Y., Bastien, D.A., Espejel, H.E., 2010. Determination of equilibrium free energy from nonequilibrium work measurements. Physical Chemistry Chemical Physics 12, 6579-6582.
de Groot, B.L., Grubmüller, H., 2001. Water Permeation Across Biological Membranes: Mechanism and Dynamics of Aquaporin-1 and GlpF. Science 294, 2353-2357.
Hansen, M., Kun, J.F., Schultz, J.E., Beitz, E., 2002. A single, bi-functional aquaglyceroporin in blood-stage Plasmodium falciparum malaria parasites. J. Biol. Chem. 277, 4874-4882.
Heymann, J.B., Engel, A., 1999. Aquaporins: Phylogeny, Structure, and Physiology of Water Channels. Physiology 14, 187-193.
Hub, J.S., de Groot, B.L., 2008. Mechanism of selectivity in aquaporins and aquaglyceroporins. Proc Natl Acad Sci U S A 105, 1198-1203.
Humphrey, W., Dalke, A., Schulten, K., 1996. VMD: Visual molecular dynamics. Journal of Molecular Graphics 14, 33-38.
Jensen, M.Ø., Mouritsen, O.G., 2006. Single-Channel Water Permeabilities of Escherichia coli Aquaporins AqpZ and GlpF. Biophysical journal 90, 2270-2284.
Lee, J.K., Khademi, S., Harries, W., Savage, D., Miercke, L., Stroud, R.M., 2004. Water and glycerol permeation through the glycerol channel GlpF and the aquaporin family. Journal of Synchrotron Radiation 11, 86-88.

MacKerell, A.D., Banavali, N., Foloppe, N., 2000. Development and current status of the CHARMM force field for nucleic acids. Biopolymers 56, 257-265.

Newby, Z.E.R., O'Connell Iii, J., Robles-Colmenares, Y., Khademi, S., Miercke, L.J., Stroud, R.M., 2008. Crystal structure of the aquaglyceroporin PfAQP from the malarial parasite Plasmodium falciparum. Nat Struct Mol Biol 15, 619-625.

Phillips, J.C., Braun, R., Wang, W., Gumbart, J., Tajkhorshid, E., Villa, E., Chipot, C., Skeel, R.D., Kalé, L., Schulten, K., 2005. Scalable molecular dynamics with NAMD. Journal of Computational Chemistry 26, 1781-1802.

Pohl, P., Saparov, S.M., Borgnia, M.J., Agre, P., 2001. Highly selective water channel activity measured by voltage clamp: analysis of planar lipid bilayers reconstituted with purified AqpZ. Proc Natl Acad Sci U S A 98, 9624-9629.

Schnick, C., Polley, S.D., Fivelman, Q.L., Ranford-Cartwright, L.C., Wilkinson, S.R., Brannigan, J.A., Wilkinson, A.J., Baker, D.A., 2009. Structure and non-essential function of glycerol kinase in Plasmodium falciparum blood stages. Molecular Microbiology 71, 533-545.

Serdyuk, I.N., Zaccai, Nathan R., and Zaccai, Joseph, 2007. Methods in Molecular Biophysics Structure, Dynamics, Function Cambridge University Press, Cambridge.

Smart, O.S., Goodfellow, J.M., Wallace, B.A., 1993. The pore dimensions of gramicidin A. Biophys. J. 65, 2455-2460.

Song, J., Almasalmeh, A., Krenc, D., Beitz, E., 2012. Molar concentrations of sorbitol and polyethylene glycol inhibit the Plasmodium aquaglyceroporin but not that of E. coli: Involvement of the channel vestibules. Biochimica et Biophysica Acta (BBA) - Biomembranes 1818, 1218-1224.

Zeidel, M.L., Ambudkar, S.V., Smith, B.L., Agre, P., 1992. Reconstitution of functional water channels in liposomes containing purified red cell CHIP28 protein. Biochemistry 31, 7436-7440.

Zhu, F., Tajkhorshid, E., Schulten, K., 2004. Collective diffusion model for water permeation through microscopic channels. Phys Rev Lett 93, 224501.

**Figures**

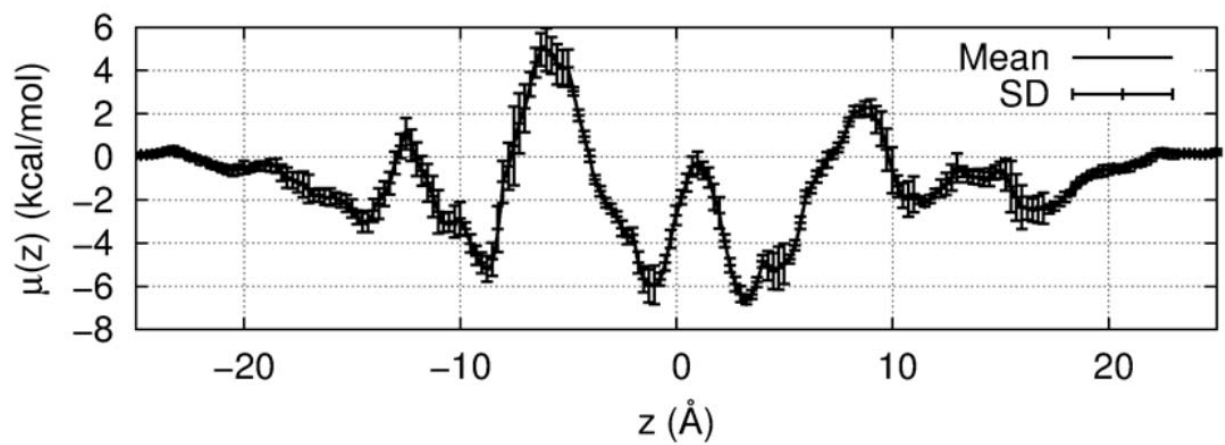

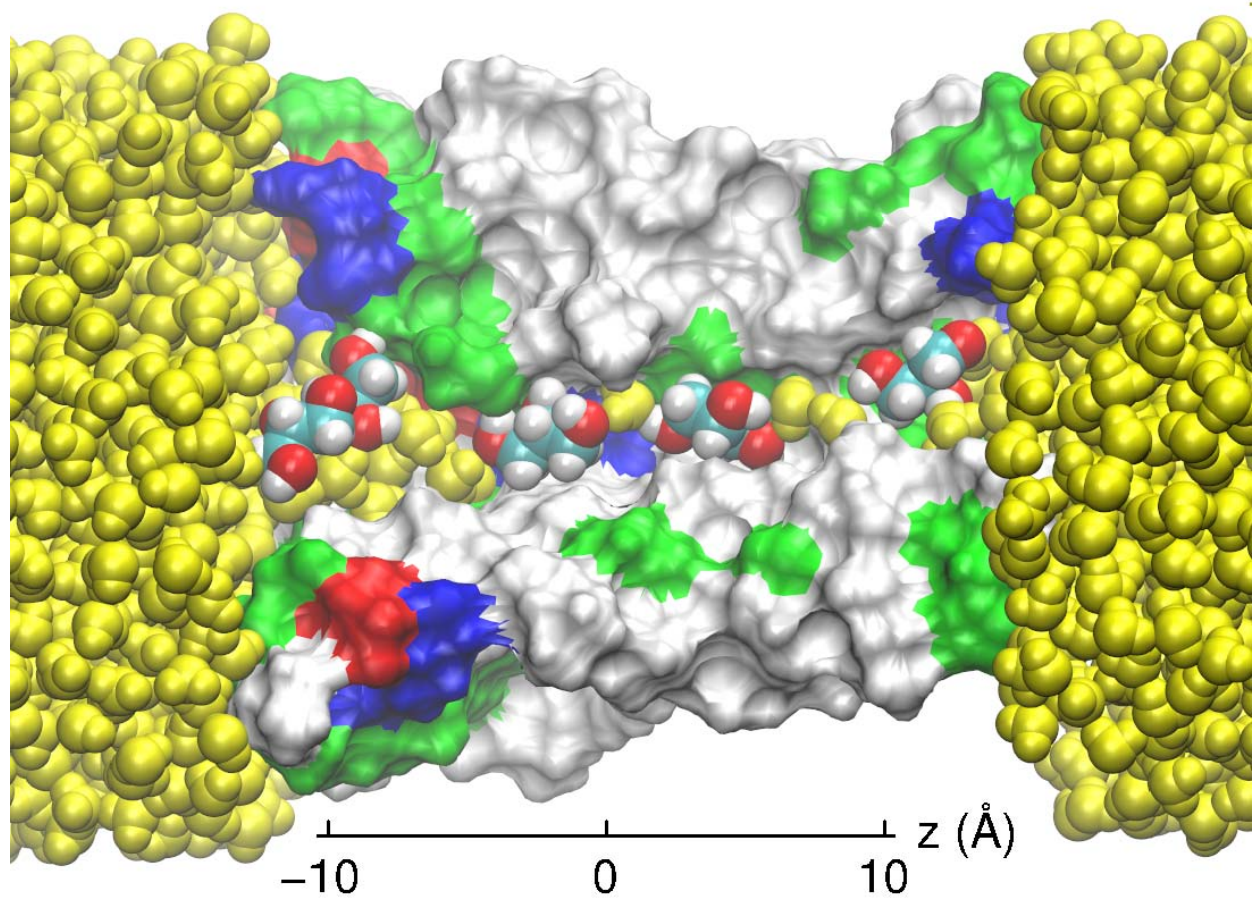

Fig. 1 Top panel, the chemical potential of glycerol though the PfAQP conducting pore *vs* its center-of-mass z-coordinate (The z-axis is normal to the cell membrane and points from the extra-cellular to the intra-cellular side). Bottom panel, binding sites of glycerol in and outside the conducting channel of PfAQP. Shown here are some of the waters (in vdw representation, colored in yellow), five glycerols (in vdw, colored by element names), and some of the protein residues (in surface, colored by residue types). Not all protein residues are shown for the purpose that the conducting channel and glycerols are easily visible. The coordinates of all the atoms are of the end state of a 3.0 ns equilibration of SysI during which all non-hydrogen atoms of the glycerols and the protein are fixed to their crystal positions (PDB code: 3C02). Graphics rendered with VMD.(Humphrey et al., 1996)

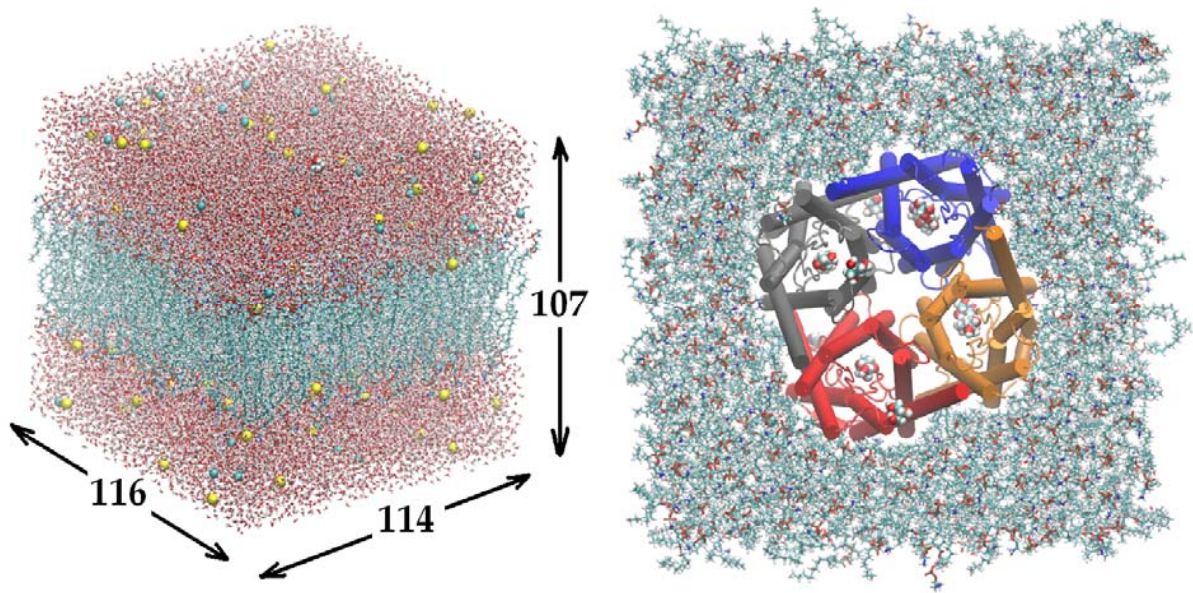

Fig. 2. All-atom model of PfAQP in the cell membrane. The system is 116Å×114 Å×107 Å in dimension. Visible in the left panel are waters (in licorice representation), lipids (licorice), ions (vdw), and one glycerol (vdw). Shown in the right panel are the PfAQP tetramer (in cartoon representation, colored by segname), lipids (licorice), and glycerols (vdw). All except PfAQP are colored by element name. Graphics rendered with VMD (Humphrey et al., 1996).

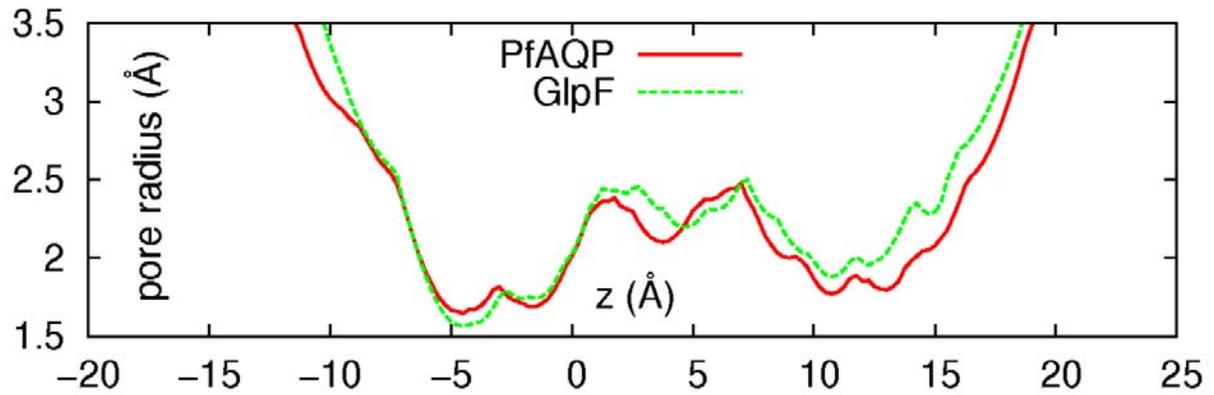

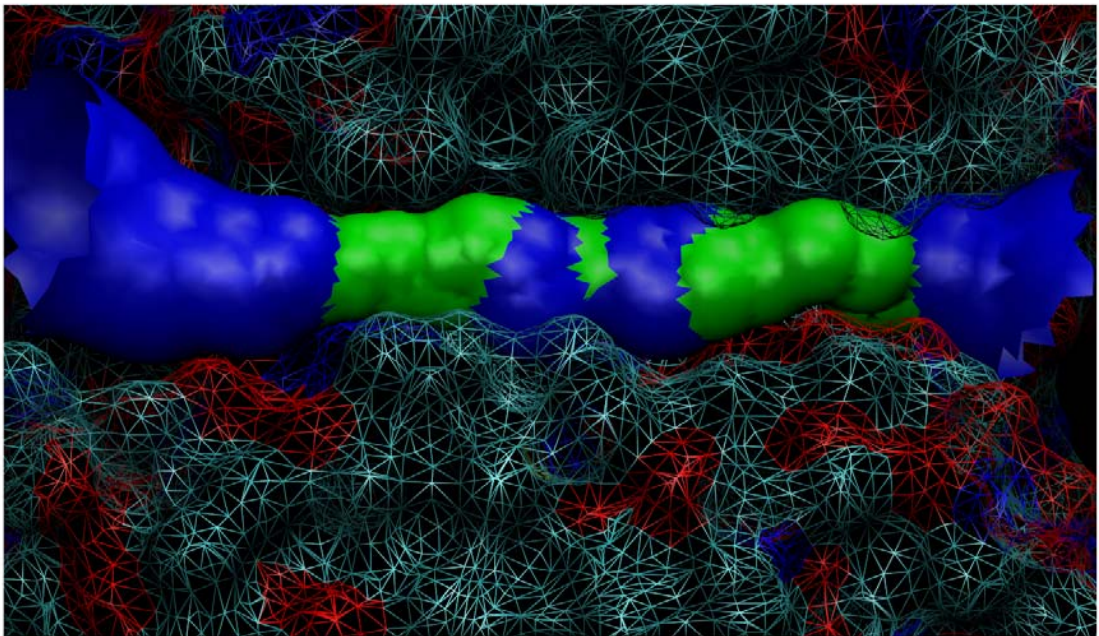

Fig. 3. Top panel, pore radius of the PfAQP conducting channel in comparison with GlpF. The radii were computed from the crystal structures (PDB codes: 3C02 for PfAQP and 1FX8 for GlpF) using HOLE2.(Smart et al., 1993) Bottom panel, illustration of the conducting pore (colored in green (narrow) and blue (wide)) along with some of the protein atoms (in wireframes, colored by element name). Not all atoms of the protein are shown for the purpose that the conducting pore is easily visible. Graphics rendered with VMD.(Humphrey et al., 1996)

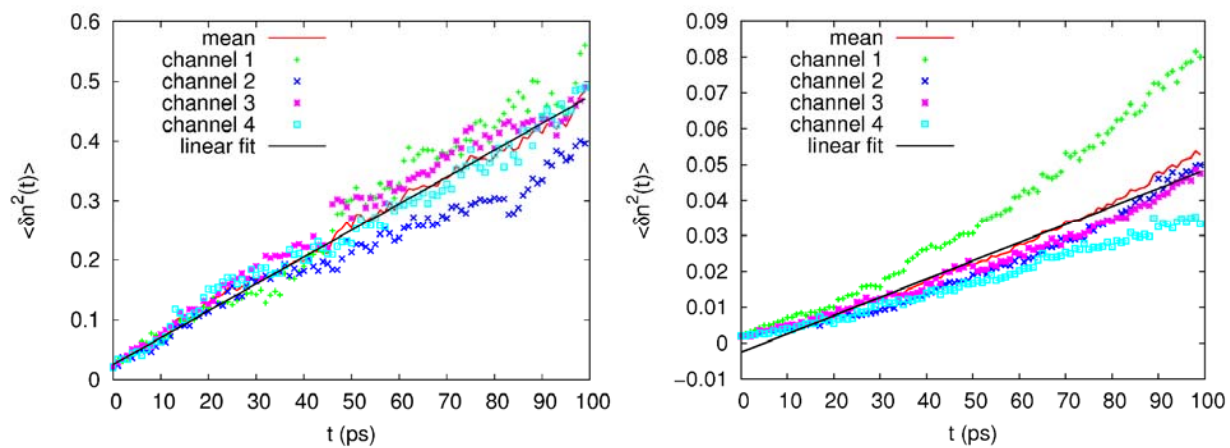

Fig. 4. The mean square displacement (msd) of waters in the conducting pores of PfAQP of SysII (left panel) and SysI (right panel). The MD trajectory was divided into segments of 0.1 ns each, over which the displacements of the waters in the pore were included in the relevant statistics.

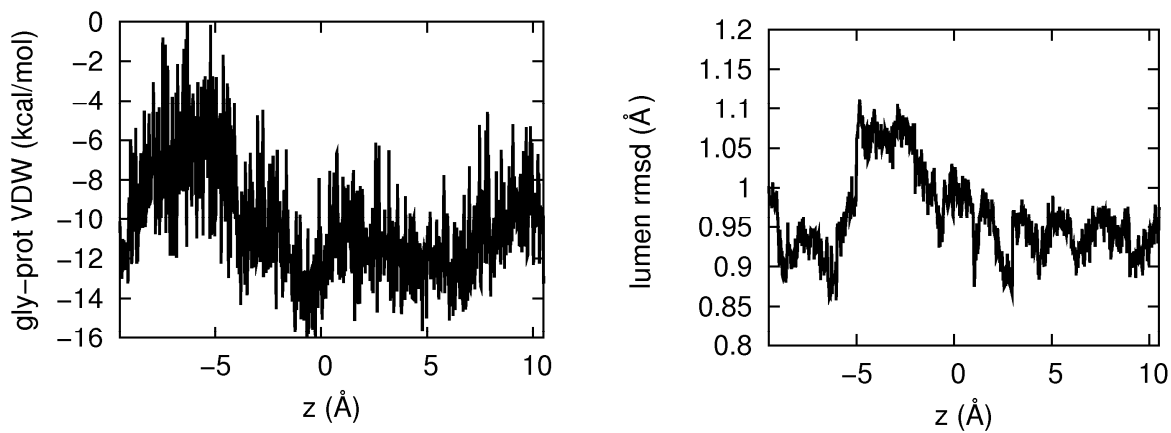

Fig. 5. Left panel, the van der Waals interaction energy between a glycerol and the protein vs the center-of-mass z-coordinate of the glycerol. Right panel, the root mean square displacement of the lumen residues of PfAQP vs the center-of-mass z-coordinate of a glycerol as it traverses the conducting channel.